# Dirac spin gapless semiconductors: Ideal platforms for massless and dissipationless spintronics and new (quantum) anomalous spin Hall effects

Xiao-Lin Wang*

*Spintronic and Electronic Materials Group, Institute for Superconducting and Electronic Materials, Australian Institute for Innovative Materials, University of Wollongong, North Wollongong, NSW 2500, Australia*

It is proposed that the new generation of spintronics should be ideally massless and dissipationless for the realization of ultra-fast and ultra-low-power spintronic devices. We demonstrate that the spin-gapless materials with linear energy dispersion are unique materials that can realize these *massless* and *dissipationless states*. Furthermore, we propose four new types of spin Hall effects which consist of spin accumulation of equal numbers of electrons and holes having the same or opposite spin polarization at the sample edge in Hall effect measurements, but with vanishing Hall voltage. These new Hall effects can be classified as (quantum) anomalous spin Hall effects. The physics for massless and dissipationless spintronics and the new spin Hall effects are presented for spin-gapless semiconductors with either linear or parabolic dispersion. New possible candidates for Dirac-type or parabolic type spin-gapless semiconductors are demonstrated in ferromagnetic monolayers of simple oxides with either honeycomb or square lattices.

Spin, charge, and mass are the electron's three attributes. Conventional electronics and information technology are based on the electron's charge and its degrees of freedom, while *spintronics* deals with the degrees of freedom of both the spin and the charge of electrons. The ultimate goals for next-generation spintronic or electronic devices should be to make them ultra-fast and ultra-low-power. This requires fast and dissipationless transport and the manipulation of both the charge and the spin of electrons at room temperature without using external conditions. To meet these requirements, it is ideal to eliminate the (effective) mass of electrons or holes and to make the massless charges fully spin-polarized.

Thanks to the Dirac dispersion, the electron's (effective) mass can be eliminated. . In 2008 the concept of the spin gapless semiconductor (SGS) [1], in which both electrons and holes can be fully spin polarized, was proposed based on the novel designs of band structures with either linear or parabolic energy dispersion. These spin-gapless semiconductors are regarded as a new class of materials, which bridge semiconductors and half-metals. Generally speaking, one spin channel has a zero gap and the other has a gap. The energy dispersion can be either parabolic or Dirac-like linear, as shown in Fig. 1. Here, we propose that the spin gapless semiconductors with the linear dispersion (Fig. 1) are unique materials that are suitable for realizing *massless spintronics* and *dissipationless spintronics*. Furthermore, we propose new types of spin Hall effects (SHE), with equal numbers of electrons and holes having the same or opposite spin polarization at the sample edge in Hall effect measurements, but with vanishing Hall voltage. We collectively denote them as the (quantum) anomalous spin Hall effect (ASHE) and quantum anomalous spin Hall effect (QASHE) for quantized cases. We discuss the details of the physics for the massless and/or dissipationless states, as well as the QASHE, for SGSs with either linear or parabolic energy dispersion.

Let us first highlight the unique band structures of the SGSs and discuss their unique spin, charge, and (effective) mass states before addressing the physics of massless and dissipationless charge transport with full-spin polarization, and their ASHE or QASHE, as well as searching for suitable materials.

Mathematically, there are four different types of zero band structures that are unique for SGSs with either linear or parabolic dispersion: Type I: zero gap for the same spin channel (spin-up in Fig. 1a); Type II: zero gap for the spin-up valence band and the spin-down conduction band (Fig. 1b); Type III: zero gap for the spin-up valence band (Fig. 1c), but without spin polarization for the conduction band; Type IV: zero gap with full spin polarization for the conduction band and zero polarization for the valence band (Fig. 1d).

Notably, all these band structures share the same features: massless states and high mobility, zero gap, and full spin polarization for both bands or one of the bands, making them ideal for *massless spintronics* or *Dirac spintronics*.

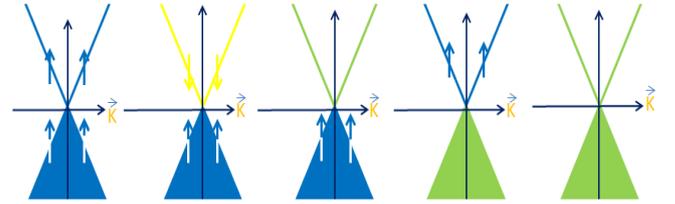

Fig. 1. Band structures of Dirac spin-gapless semiconductors with linear dispersion for types I (a), II (b), III (c), and IV (d). (e) is the band structure for spin-degenerate Dirac systems.

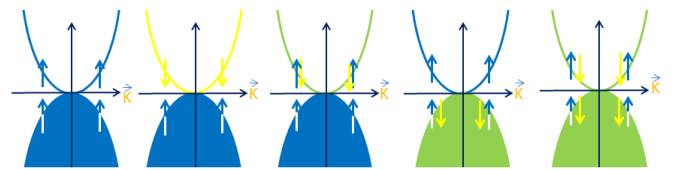

Fig. 2. Band structures of spin-gapless semiconductors with parabolic dispersion for types I (a), II (b), III (c), and IV (d). (e) is the band structure for spin-degenerate systems.

Beside the massless states with full spin polarization, the Dirac spin-gapless semiconductors exhibit more exotic spin and charge states. Gating or chemical doping can induce fully polarized electrons or holes, or excitons with the same spin sign in type I, spin-up holes or spin-down electrons in type II, spin-up holes in type III, and spin-up electrons in type IV, respectively [1, 2]. Under light or thermal excitation, equal numbers of both excited holes and electrons are fully spin polarized [1] in type I, as well as excited holes and electrons with opposite spin for type II, fully polarized holes and non-polarized electrons in type III, and fully polarized electrons and non-polarized holes in type IV. It should be pointed out that the Dirac type dispersion gives rise to intrinsically zero-gap band structures with the Fermi level exactly crossing the Dirac point for high symmetry structures such as honeycomb lattices. Therefore, little energy is needed to realize these excitation states. This makes charge concentrations easily tuneable by gating, doping, or excitation in SGSs, in contrast to half-metals, which usually have high charge concentrations.

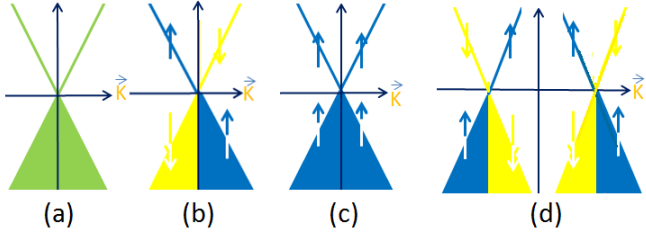

Fig. 3. Band structures of Dirac-like systems: spin degenerate systems (a), the surface or edge states of topological insulators (b), spin-gapless state of type I (c), and Weyl semimetals (d).

The band structure of the type I SGSs with linear dispersion is compared to that of other Dirac-type systems (Fig. 3): Dirac-type zero-gap systems such as graphene and three-dimensional (3D) Dirac systems (Fig. 3a), and edge or surface states in two-dimensional (2D) or 3D topological insulators (Fig. 3b), as well as Weyl semimetals [2-8] (Fig. 3d). There is no spin polarization in graphene and Weyl metals, or spin is not fully polarized, but it exhibits spin-momentum locking in topological insulators (TIs). This is in great contrast to the type I SGSs, in which the spin is fully polarized for both electrons and holes. We have now seen that SGSs with (Dirac-like) linear dispersion are ideal platforms for the realization of *massless spintronics with full spin polarization*.

We now discuss how to separate fully polarized charges and achieve dissipationless charge transport in SGSs with either linear or parabolic dispersion. It has been proposed [1] that by using the Hall effect (through either an external field or internal magnetization), fully spin polarized electrons or holes can be easily separated and accumulated at sample edges, with this mechanism denoted as the spin filter effect [1].

We first discuss the various Hall effects under internal magnetization for the case of ferromagnetic SGSs with the Fermi level ($E_F$) level penetrating into either the conduction or the valence band. For convenience in discussion, Fig. 4 provides schematic illustrations of all the existing Hall effects, such as Hall effect (HE) (Fig. 4a), the anomalous Hall effect (AHE) (Fig. 4b), the spin Hall effect (SHE) [9,10] (Fig. 4c), the quantum Hall effect (QHE) (Fig. 4d), the quantum anomalous Hall effect (QAHE) [11,12] (Fig. 4e), and the quantum spin Hall effect (Fig. 4f) [5] , respectively. With gating or doping (which slightly shifts $E_F$ up or down), charge deflection to the sample edge comes from 1) spin-up electrons or holes in type I (AHE); 2) spin-up holes or spin-down electrons in type II (AHE); 3) non-polarized electrons (HE) or spin-up holes (AHE) in type III; and 4) spin-up electrons (AHE) or non-spin-polarized holes (HE) in type IV, respectively. Note that the Hall effects in Dirac SGSs involve massless states for spin-polarized charges or the non-polarized cases.

Now, we propose four new types of *spin Hall effect* as a result of the unique band structures of SGSs with both linear and parabolic dispersions due to either internal magnetization or an external magnetic field under light or thermal excitation.

Light or thermal excitations can produce equal concentrations of both electrons and hole, with the electrons and holes having the same spin sign in type I SGSs, opposite spin in type II, spin-up for holes and non-spin polarized electrons in type III, and spin-up electrons and non-spin polarized holes in type IV, respectively. Remarkably, the internal magnetization can produce four new types of (quantum) spin Hall effects with vanishing Hall voltage ($V_H$ = 0), depending on which electrons and holes deflect to and accumulate at the same edge: 1) spin-up electrons and spin-up holes (Fig. 5a) for type I; 2) spin-down holes and spin-up electrons for type II (Fig. 5b); 3) spin-down holes and non-spin-polarized electrons for type III (Fig. 5c); and 4) spin-up electrons and non-spin polarized holes for type IV (Fig. 5d), respectively.

The vanishing Hall conductance for these effects is the same as what is seen in SHE and QSHE. These new types of spin accumulation are different from the spin Hall effect, however, in which spin-up and spin-down electrons accumulate at two opposite edges as a result of an *electric field,* and differs from QSH, with spin-up and spin-down electrons moving in opposite directions along edges due to the time reversal symmetry protection predicted for 2D TIs [5]. We classify the new types of spin Hall effect as belonging to the *anomalous spin Hall effect (ASHE)* or the *quantum ASHE (QASHE)* (Fig. 5e-g) for quantized cases which resemble the AHE and QAHE.

Let us now look into how the dissipationless state is intrinsically possible in SGSs without using gating or doping, i.e., the $E_F$ level is locates at the Dirac point or at the point where the bottom of the conduction band touches the top of the valence band for the parabolic case. Naturally, we should consider the quantum anomalous Hall effect proposed in magnetically doped TIs [11] due to spin-orbit coupling or charge and internal magnetization interaction. If the SGSs are ferromagnetic, the internal magnetization, as well as the strong spin-orbit coupling of either magnetic transition metals or rare earth magnetic elements, will open a gap (or localize free charges internally), which, in turn, leads to dissipationless edge states thanks to the principles of the QAHE [11]. Obviously, dissipationless transport is attainable when the internal magnetization is high enough in type I SGSs (having the same spin direction for both bands) with either linear or parabolic dispersion due to the non-trivial insulating state and the breaking of time reversal symmetry for the edge state [11]. It should be noted that the dissipationless edge state also works the same way for parabolic type I SGSs. The dissipationless edge state is absent, however, for other types of SGSs. At the Fermi level, one occupied spin channel lowers its energy by magnetization, but the other occupied spin channel stays in the conduction band, leaving the system metallic. *This is very important for guiding the search for ideal candidates for the realization of room temperature (RT) QAHE (RT dissipationless edge transport without external field), ferromagnetic Dirac SGSs with the same spin direction, and Curie temperatures above RT.*

So far, the magnetically doped TIs are the candidates for the QAHE. It was only observed in them at extremely low temperatures, however, due to the very large effective mass (parabolic dispersion) and low Curie temperature [12]. The ferromagnetic SGSs with linear dispersion, easily tuneable charge concentration, and high Curie temperatures are clearly ideal platforms for easy realization of the QAHE at higher temperatures, thanks to the high mobility of their charge carriers as a result of massless states, full spin polarization, and small charge density. These are the advantages of Dirac SGSs over the parabolic dispersion, which gives massive charges with higher charge concentrations, leading to low mobility.

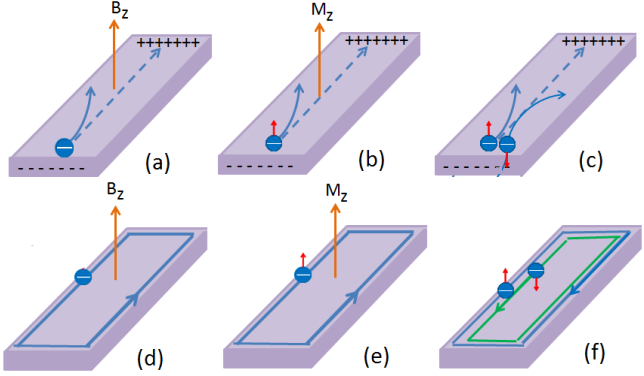

Fig. 4. Schematic illustrations of the Hall effect (a), anomalous Hall effect (b), spin Hall effect (c), quantum Hall effect (d), quantum anomalous Hall effect, and quantum spin Hall effect (f).

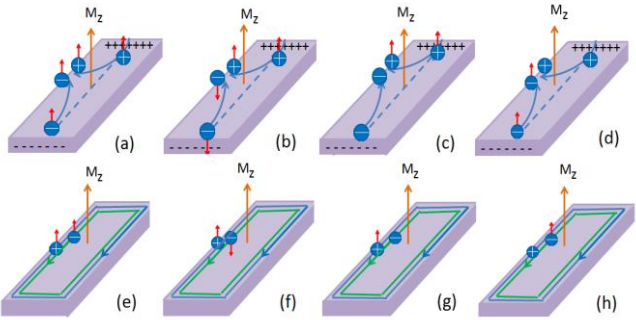

Fig. 5. Schematic illustrations of novel spin Hall effects with vanishing Hall voltage, namely anomalous spin Hall effects (upper panel) and quantum anomalous spin Hall effects (lower panel) for spin-gapless semiconductors with either linear (Dirac-type) or parabolic type I (a, e), II (b, f), III (c, g), and IV (d, h) behaviour. "+" and "-" denote holes and electrons, respectively. Note that the electrons and holes have same spin for (a, e) but opposite spins for (b, f). All cases have vanishing Hall voltage.

We now come to the issue of suitable real materials which can offer massless, dissipationless states and the new types of spin Hall effects. Since the SGSs were proposed and verified for one of the four types (type II with parabolic dispersion) in oxide in 2008, great attention and interest have been attracted to the search for more suitable candidates both experimentally and theoretically [12-20]. It has been proposed that the spin-gapless features are possible in a wide range of gapless and narrow-gap oxides, non-oxide semiconductors, ferromagnetic and antiferromagnetic semiconductors, and non-oxides with appropriate elemental substitutions [1,2]. Since then, SGSs with parabolic dispersion have been verified, mostly theoretically, in many different systems [1,2, 12-20].

The first experimental realization of the SGSs (type I with parabolic dispersion) was successfully achieved in the Heusler compound $Mn_2CoAl$ [14]. An anomalous small Hall conductance and zero Seebeck coefficient were observed. Note that a zero-gap material, HgTe doped with Mn, was predicted to show the QAHE [5]. It should be noted that a typical spin-gapless state having the same spin direction for both the valence and the conduction band (the same as in the type I SGSs) is the critical transition point before band inversion occurs, which leads to the QAHE due to spin-orbit coupling. That is to say, the spin-gapless state can be regarded as the precursor for gap formation by the internal charges responsible for the dissipationless edge state.

As pointed out earlier, massless spintronics requires SGSs with linear dispersion or Dirac SGSs. Also, the room temperature dissipationless state should be achievable relatively easily in the Dirac type I (massless) case compared to the parabolic type I (massive) case, so long as the Curie temperature is higher than room temperature. Any suitable graphene (Dirac) type materials with chemical substitutions were suggested to be good candidates due to their Dirac-like dispersion [1,2] or honeycomb lattice structures. It has come to our attention that a honeycomb ferromagnetic lattice [21] shows the same features as in Fig. 1a, which verifies the SGSs with linear dispersion that we proposed in 2008. The type I SGSs with linear or parabolic dispersion have been successfully verified in a honeycomb lattice of modified tri-s-triazine ($C_7N_6$) units with fully spin-polarized Dirac cones and in a hybrid honeycomb lattice composed of $C_7N_6$ and s-triazine ($C_3N_3$) units with parabolic dispersion, with ferromagnetic ordering and Curie temperature of 830 and 205 K for the two lattices, respectively [19]. The two honeycomb lattices have topologically nontrivial electronic states suggesting that quantum anomalous Hall effect (QAHE) states could be achieved in these metal-free materials.

A few other works have also come to our attention that demonstrate such band structures in organic systems with transition metals arranged in honeycomb structures [22]. They also verify the spin-gapless band structures that we proposed for the SGSs with linear dispersions (Type I, Fig. 1a)). Very recently, band structures showing the exact features of Fig. 1a were demonstrated in Mn-intercalated epitaxial graphene [23]. The QAHE was also predicted in this system. There is another very recent report on Cr-interacted $TiO_2$, which also shows the same band structure as the SGSs with linear dispersion [24].

All these reports, in fact, have verified the proposed band structure of the type I Dirac SGSs, and they all predict the QAHE with dissipationless edge states. It will be a great challenge to fabricate real samples based on these candidates, however, due to technical difficulties. From the perspective of applications, we believe that simple oxides should be ideal candidates, as they are stable in the ambient environment. There two key selection criteria for the candidates for massless and/or dissipationless spintronics and ASHE: 1) ferromagnetic; 2) suitable lattices that give rise to Dirac-type (linear dispersion) band structures; 3) For the parabolic case, the zero gap should be direct.

Here, we demonstrate that spin-gapless band structures with linear or parabolic dispersion can be further demonstrated in monolayers of a simple oxide such as MnO or $VO_2$ with a honeycomb or square lattice, respectively. There are many materials having honeycomb lattices. The typical example is ZnO, which crystallizes in a hexagonal structure. Remarkably, its surfaces are atomically flat, stable, and exhibit no reconstruction. Therefore, it could provide the perfect honeycomb lattice surface for epitaxial growth of monolayers of other simple magnetic or non-magnetic oxides. We have performed calculations on the band structure of a ferromagnetic monolayer of MnO in a hexagonal lattice. Geometry optimization was done before carrying out the calculations on the strain effect on the tuning of band structures.

After geometry optimization, we found that the buckling of the monolayer MnO disappears and that it becomes extremely flat with a perfect honeycomb lattice (Fig. 6a). Fig. 6 shows the band structures having parabolic (b, c) or linear dispersion (c) for different strain effects or lattices. The bottoms of the conduction bands of both spin-up and spin-down electrons overlap and touch the top of the spin-up valence band exactly at $E_F$ with zero gap, corresponding to the type III SGSs (Fig. 2c). On the other hand, slight tensile strain produces the type I band structure with zero gap for both conduction and valence band carriers with the same spin direction (Fig. 2a). Under further tensile strain, Dirac type conduction and valence bands with the same spin direction touch at the Dirac point or Fermi level, as shown in Fig. 6d. The enlarged details of the band structures show that the dispersion is perfectly linear, and the bands retain the zero gap with the Dirac point crossing the $E_F$.

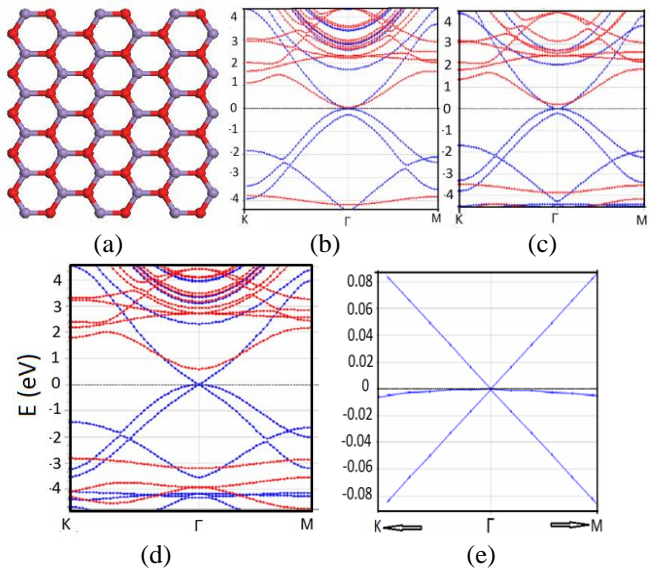

Fig.6. MnO monolayer in honeycomb lattice (a), and its band structures for different lattices: $a$ = 3.40 Å (b), $a$ = 3.42 Å (c), and $a$ = 3.50 Å (d). (e) is an enlargement of the details for the bands near $E_F$ in (d). Blue and red curves represent spin-up and spin-down, respectively.

We further demonstrate the linear type of spin-polarized zero-gap band structure in a ferromagnetic $VO_2$ monolayer with a square lattice (Fig. 7a). A perfect linear dispersion at $E_F$ with the same spin direction for both the conduction and the valence band is revealed in Fig. 7b and c. We have also obtained the same type I behavior with parabolic dispersion in ferromagnetic monolayer CoO.

It is very likely that the ASHE or QASHE can be achieved in a ferromagnetic monolayer of MnO or $VO_2$, as both the conduction and the valence bands are direct bands and almost symmetrical. Both massless and dissipationless states with fully spin-polarized charge are intrinsically possible for the cases of Fig. 6d for MnO, and Fig. 7 for $VO_2$, but only dissipationless states are possible for their parabolic cases.

It should be noted that if the candidate is non-magnetic, the ferromagnetic state is attainable using the magnetic proximity effect, as demonstrated by the very recent experimental observation of a high temperature ferromagnetic topological insulating state in $Bi_2Se_3$ grown on ferromagnetic EuO [].

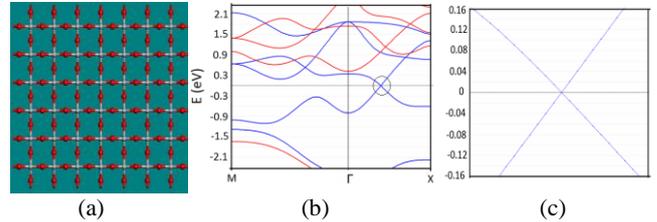

Fig. 7. $VO_2$ monolayer with square lattice (a), and its band structure (b). (c) shows the enlarged details for the bands near $E_F$ in (b). Blue and red curves represent spin-up and spin-down, respectively.

We have demonstrated that the spin-gapless materials with linear energy dispersion are unique materials for realizing *massless* and *dissipationless states*. The dissipationless states can also be achieved in SGSs with parabolic dispersion. Both types of SGSs can lead to (quantum) anomalous spin Hall effects, which consist of spin accumulation of equal numbers of electrons and holes having the same or opposite spin polarization at a sample edge during Hall effect measurements, but with vanishing Hall voltage. We have further demonstrated that it is very likely that SGSs with either linear or parabolic dispersion can commonly exist in the form of monolayers of either honeycomb or square lattices for a wide range of other transition metals or rare earth oxides in ferromagnetic states. The fascination of massless and dissipationless spintronics as well as the proposed novel spin Hall effects is likely to stimulate more interest in searching for more candidates for 2D [25] or even 3D SGSs. Experimental realizations of these states at both low temperature and RT are imminent.

Acknowledgements

The author is grateful for funding support from the Australian Research Council for this work and thanks Dr. Tania Silver for polishing the English of this manuscript.

*Email: xiaolin@uow.edu.au